\documentclass{article}
\usepackage{spconf,amsmath,graphicx}
\usepackage{algorithm}
\usepackage{algpseudocode}
\usepackage{diagbox}
\usepackage{amssymb}
\usepackage{adjustbox}
\usepackage{array}
\usepackage{hyperref}

\title{Combinatorial Music generation model with song structure graph analysis}

\twoauthors
  {Seonghyeon Go}
	{M.S. Graduate, \\
 Department of Intelligence and Information\\
 Seoul National University\\}
  {\hspace{-15mm}Kyogu Lee}
{\hspace{-15mm}Department of Intelligence and Information\\
\hspace{-15mm}Interdisciplinary Program in Artificial Intelligence\\
\hspace{-15mm}AI Institute, Seoul National University}
\begin{document}
\ninept
\maketitle
\begin{abstract}
In this work, we propose a symbolic music generation model with the song structure graph analysis network. We construct a graph that uses information such as note sequence and instrument as node features, while the correlation between note sequences acts as the edge feature. We trained a Graph Neural Network to obtain node representation in the graph, then we use node representation as input of Unet to generate CONLON pianoroll image latent. The outcomes of our experimental results show that the proposed model can generate a comprehensive form of music. Our approach represents a promising and innovative method for symbolic music generation and holds potential applications in various fields in Music Information Retreival, including music composition, music classification, and music inpainting systems. 

\end{abstract}
\begin{keywords}
combinatorial music generation, graph neural network, song structure analysis
\end{keywords}
\section{Introduction}\label{sec:introduction}

Recent research in the field of symbolic music generation has been focused on utilizing deep learning-based generation models like GAN\cite{dong2018musegan} and Transformer\cite{huang2018music} to produce high-quality music. However, several challenges remain for music-generating models. For example, the ability to consider the structure of music during the generation process, and the creation of commercially viable music still poses a significant challenge.

Theme transformer\cite{shih2022theme} model has been used to maintain a regular song structure by directly conditioning the melody used as the main theme. Another model called MELONS\cite{zou2022melons} has divided songs into nodes to form a graph and conducted a study on generating music by analyzing the relationship between these nodes. These models were suggested to consider the song structure, but they do not consider the song structure in multi-track case.

Lee et al.\cite{lee2022commu} proposed the concept of combinatorial music generation to increase the commercial applicability of music generation models. In stage 1, we should generate note sequences containing a specific role, like instrument, track-role, or genre. In stage 2, the note sequences are combined to create complete music. This form of music generation resembles the form of music produced by real people, and it is easy to commercialize the generated results.

\begin{figure}[t]
 \centerline{
 \includegraphics[width=0.9\columnwidth]{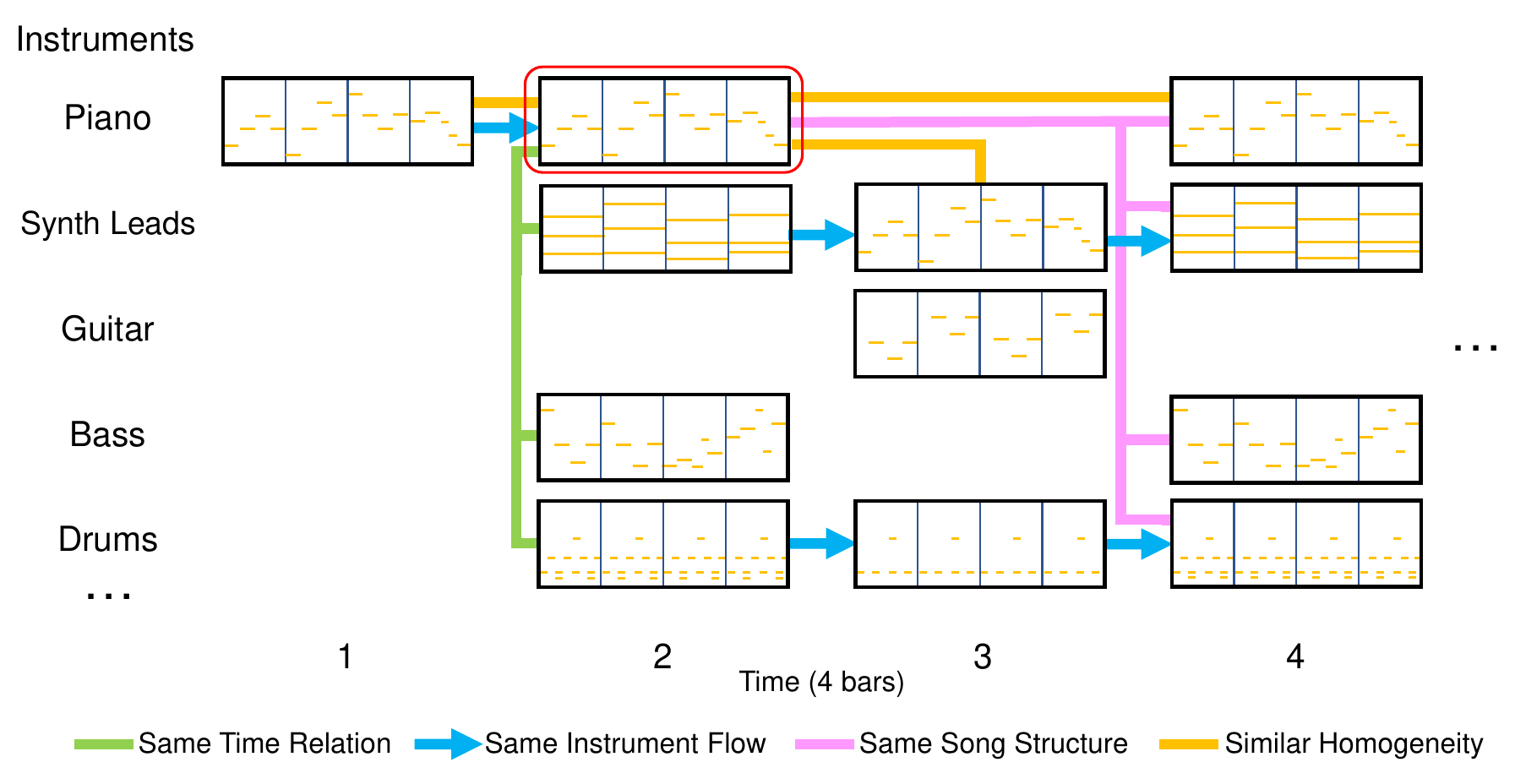}}
 \caption{Song Structure Graph example. There are edges related to a musical pattern in the red box, except for some "Same Instrument Flow" edges. Note that this figure plots general pianoroll, not CONLON pianoroll\cite{angioloni2020conlon} which used in our study.}
 \label{fig:graph}
\end{figure}

To address these challenges, we introduce a new framework that analyzes song structure using graphs for combinatorial music generation. We analyzed the given music's song structure based on the simplicity matrix to construct a song structure graph. Example of song structure graph is illustrated in Figure \ref{fig:graph}. Each node in the graph represents pieces of information containing metadata of bars, such as note sequences, keys, and instruments. We denote these pieces of information as "Musical Pattern" in this paper. Each edge in the graph represents the relationship between musical patterns. We then use Relational Graph Neural Network to obtain the representation of each musical pattern node. This node representation is then used to generate a piano-roll image. Our proposed framework enables the understanding of the overall structure of multi-track music and facilitates the generation of each musical pattern for music generation. We also show that it can be used for symbolic music generation or inpainting by proposing a framework that can directly obtain structure structures by using graphs at the bar level.
\section{Background}

\subsection{Song Structure Analysis}

Song structure analysis based on the self-similarity matrix is a well-established method, especially in the audio domain\cite{article}. Various studies have also been conducted on song structure analysis in symbolic music\cite{lazzari2023pitchclass2vec}\cite{han2022symbolic}. However, in the symbolic music domain, this method is not robust to variations like pitch shifts. With this issue, we adopted the idea of content-based image retrieval\cite{jain2015content}. We employed a Wasserstein autoencoder\cite{tolstikhin2017wasserstein} to transform each piano-roll into the embedding of the latent space, and then we computed cosine similarity to conduct song structure analysis.

\subsection{Graph Neural Network}
Graph Neural Network (GNN) was first proposed by \cite{gori2005new}, which makes it easy to obtain a representation for each node for graph data. Recently, numerous studies have applied GNNs to Music Information Retrieval (MIR)  tasks\cite{karystinaios2022cadence}\cite{da2022heterogeneous}\cite{jeong2019graph}. In most of these studies\cite{karystinaios2022cadence}\cite{jeong2019graph}, a single note was treated as a node, and the relationship between notes was represented as an edge. This approach might provide only a limited view of the entire song structure. To address this issue, we expand the scope to include musical patterns as nodes, and each edge represents the relationship between musical patterns. In this manner, we treat each complete music as a single graph.

We used the Relational Graph Convolutional Network (RGCN) to analyze the features of various edges and nodes. For directed and labeled graphs as $G = (\mathcal{V},\mathcal{E},\mathcal{R})$ with nodes $v_{i}\in \mathcal{V}$ and edges $(v_{i},r,v_j) \in \mathcal{E}$ and relation type $r\in \mathcal{R}$, we can compute hidden representation in $(l+1)^{th}$ layer in RGCN as the following equation:
\begin{equation}
    h_{i}^{l+1} = \sigma\left( W_{0}^{(l)}h_{i}^{(l)} + \sum_{r \in \mathcal{R}}\sum_{j \in N_{i}^{r}}\frac{1}{c_{i,r}}W_{r}^{(l)}h_{j}^{(l)}\right)
\end{equation}

where $N_{i}^{r}$ denotes the set of neighbor indices of node $i$ under relation $r \in \mathcal{R}$, $c_{i,r}$ is a problem-specific normalization constant parameter. we use $c_{i,r} = |N_{i}^{r}|$ for experiment.

\section{Proposed method}\label{sec:page_size}

\subsection{Song Structure Graph}

\begin{figure}[t]
 \centerline{
 \includegraphics[width=0.9\columnwidth]{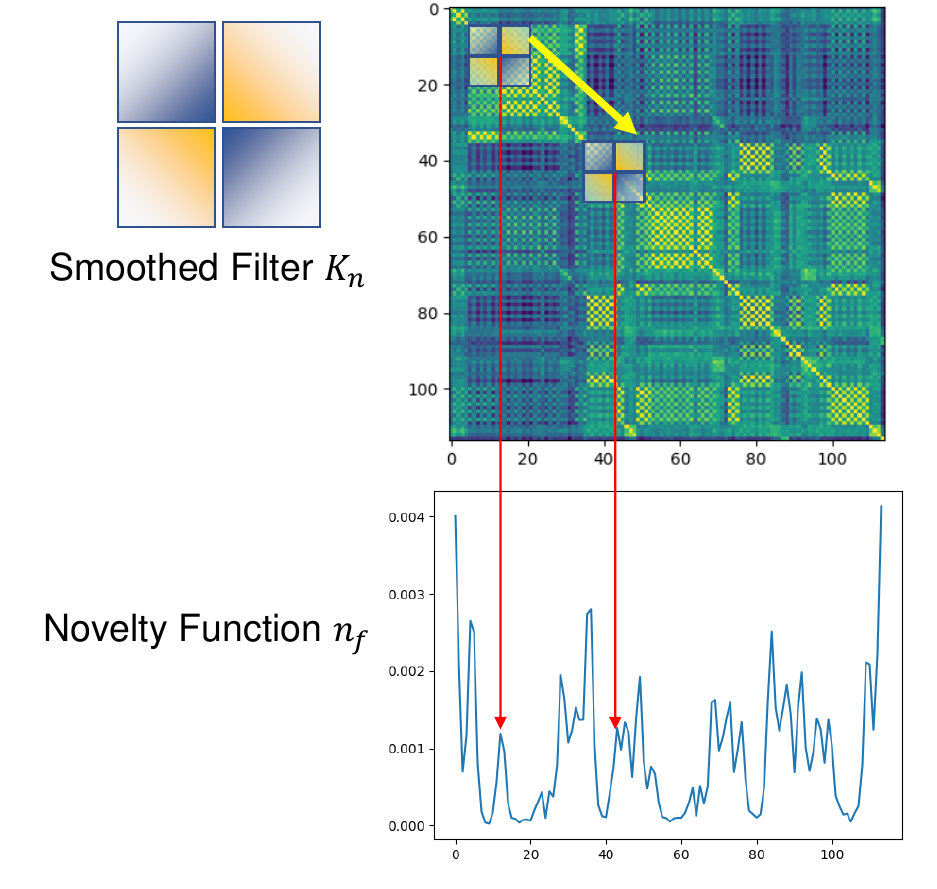}}
 \caption{Kernel function for compute novelty function $n_{f}$.}
 \label{fig:novelty}
\end{figure}

\newcommand{\factorial}{\ensuremath{\mbox{\sc Factorial}}}
\begin{algorithm}[t]
\caption{Song Structure Graph Constructing Algorithm}\label{euclid}
\begin{algorithmic}[1]
\Require{Music's MIDI data $M$}
\Require{Music's metadata $meta$}
\Function{getGraph}{$M,meta$}
\State $B_{N,I} \gets$ \Call{split}{$M$} %\Comment{Split MIDI into Bar Sequence $B$ with length $N$, \# of instrument $I$.}
\State $SM_{i,j} \gets$ \Call{SimilarityMetric}{$B_{i},B_{j}$} $\forall i,j  \in N$  % \Comment{Distance can be any metric with codomain $\in$ [0,1]}
    \For{$k$ in \Call{range}{N}}
        \State$NC_k \gets$ \Call{kernel function}{$SM, k$} % \Comment{Get Novelty Curve values for position $k$.}
    \EndFor
\State $SP \gets \{ \}$  \Comment{Starting Points for Musical Patterns}
\State $MP \gets \{ \}$  \Comment{Musical Patterns}
    \For{$l$ in \Call{range}{N}} 
        \If{$NC_l>$ threshold$\And $it is local maxima}
            \State $SP$.insert($l$) %\Comment{Set starting points as the time where local maxima of novelty curve occur.}
        \EndIf
    \EndFor
    \For{$t$ in $SP$}
        \If{$t$ is not in $Used$}  %\Comment{ex) 4 and 6 in $SP$, we can't use 6 for SP.(4$~$8)}
            \If{$B_{[t,t+pl],inst}$ is not empty} % \Comment{pl is length of patterns.}
                \State $MP.$insert$({[B,meta]_{[t,t+pl],inst}})$ 
                \State $Used\gets{[t,t+pl]}$
            \EndIf
        \EndIf
    \EndFor
\State{$V \gets MP$} % \Comment{Use Musical Patterns as nodes.} 
\State{$E \gets \{\}$} 
    \For{$u \in V$} 
        \For{$v \in V$} % \Comment{From node $u$ to node $v$}
            \If{\Call{Rel}{$u,v$}} % \Comment{Rel() function returns edge feature if there is a relation between nodes.}
                \State $E$.insert$(u,v,\Call{Rel}{u,v})$  
            \EndIf
        \EndFor
    \EndFor
\State{$g \gets G(V,E)$} %\Comment{Define song structure graph.}

\Return{$g$}
\EndFunction
\end{algorithmic}
\label{algorithm}

\end{algorithm}

\begin{figure*}[t]
 \centerline{
 \includegraphics[width=2.0\columnwidth]{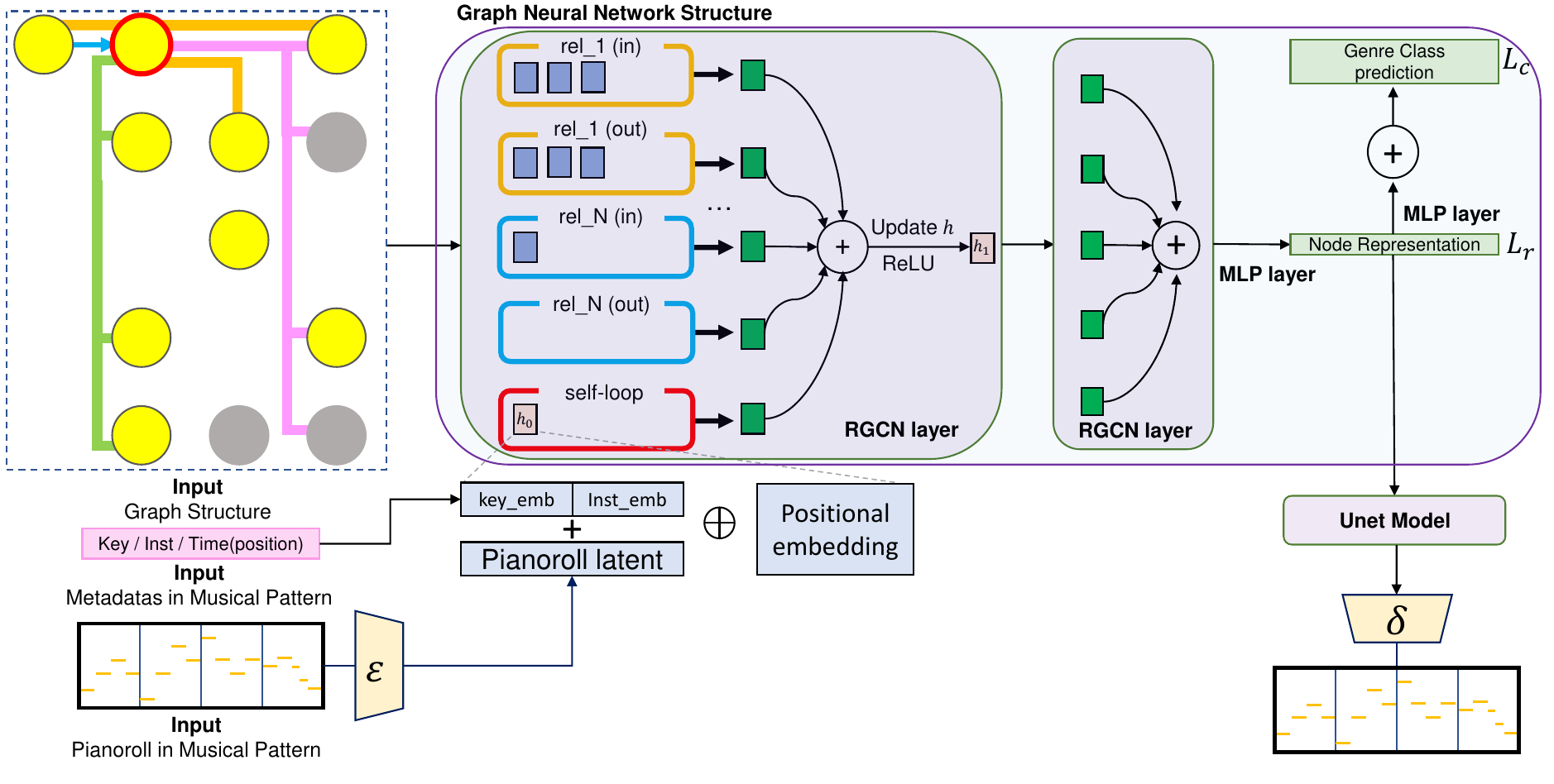}}
 \caption{Red-circled node's hidden state update in the RGCN layer. $\epsilon$ means encoder and $\delta$ means decoder.}
 \label{fig:gnn}
\end{figure*}

The construction of the song structure graph involved the application of the self-similarity matrix to the analysis of song structures. The given MIDI data was divided into bar sequence $BS$ with length $N$. The image was compressed into a Wasserstein Autoencoder (WAE) to compute the cosine similarity in latent space. The self-similarity matrix was computed by the formula:
\begin{equation}
SM_{i,j} = \frac{1}{2}+\frac{w(BS_{i})\cdot w(BS_{j})}{2\lVert w(BS_{i})\rVert \lVert w(BS_{j})\rVert},  i,j\in N 
\end{equation}
where $w$ is the encoder of the WAE. We modified some terms to make $SSM_{i,j} \in [0,1]$ for all $i,j\in N$. We can compute novelty function $n_f$ with kernel matrix $K_{N}$\cite{muller2015fundamentals}\cite{rodrigues2022feature},
\begin{equation}
K_{N}(i,j) = sign(a)\cdot sign(b),  a,b \in [-L, L]
\end{equation}
where $L$ is the range for the kernel function, and $sign$ is the sign function (-1, 0, or 1). We use the radially Gaussian function to smooth the kernel,
\begin{equation}
    \phi(a,b) = exp\left(-\frac{a^2+b^2}{2L\sigma^2}\right)
\end{equation}
where $\sigma$ is the standard deviation for $a$ and $b$. The novelty function $n_f$ can be calculated by correlating the kernel with the diagonal of the self-similarity matrix. These processes are described in Figure \ref{fig:novelty}.

\begin{equation}
    n_{f}(m) = \sum_{a,b=-L}^{L}K_{N}(a,b)\phi(a,b)SSM_{m+a, m+b} 
\end{equation}

This study employs a homogeneity and repetition-based approach to gauge the similarity of given musical patterns and to identify any recurring repetitions of these patterns at subsequent times. Using the Similarity Matrix, we can detect repetitions in music by identifying high values in the same row. We used 7 Hu moment based image similarity\cite{hu1962visual} to compute Homogeneity between two pianoroll. The resulting information is then employed to define the various edges of the song structure graph. Our whole algorithm for constructing the song structure graph is described in Algorithm \ref{algorithm}. Edges used in our experiments are explained below.

\begin{description}
    \item[Same Time Relation] Connect if two musical patterns playing at the same time.
    \item[Same Instrument Flow] Connect if the same instrument is played continuously. Note that these are direct edges.
    \item[Same Song Structure] Connect musical patterns in time $i$ and $j$, if $SSM_{i,j}>\mathbf{t}$, where $\mathbf{t}$ is threshold.
    \item[Similar Homogeneity] Connect if two musical Pattern has a similar shape based on 7 Hu invariants.
\end{description}

\subsection{Model Architecture}

In our training framework, we employed a Deep Convolutional Autoencoder with a residual block to compress the piano roll image data into the latent space. We used the Mean Squared Error (MSE) loss function to reduce reconstruction errors. We also obtained the feature vector through the embedding layer to use the key, instrument, and time information given in the musical pattern.

In the inference process of generating music, there are cases where the pianoroll feature is not given to the musical Pattern node. Based on \cite{shi2020masked}, We can get proper embedding by simply adding the "Observed" feature's and a partially observed feature in the same embedding space. Therefore, we used the sum of features that embeds key/instrument information and piano roll latent embedding. We use zero vector for masked piano-roll's latent embedding. The time information embedding was concatenated into a combined feature vector to serve as a positional embedding. Thus, the initial state $h_{0}$ can be computed as:
\begin{equation}
    h_{0} = (X + \hat{Y})\oplus T, X \in \mathbb{R}^{d_{h}}, \hat{Y} \in \mathbb{R}^{d_{h}}, T \in \mathbb{R}^{d_{t}}
\end{equation}
where $d_{h}, d_{t}$ is hidden dimension of features and time embedding, $X$ is the embedding of observed information(key/instrument), $\hat{Y}$ is embedding of partially observed information(pianoroll), and $T$ is the positional embedding for each musical patterns. Shi et al. \cite{shi2020masked} have shown that this model can be approximately decomposed into observed feature propagation and partially observed information. 
Augmentation methods such as DropEdge\cite{rong2019dropedge} and DropNode\cite{you2020graph} were used to obtain more generalized performance. And we used two additional loss functions. MSE Loss function was used to obtain the latent of the masked image, and the other was to use cross-entropy loss to proceed with the task of predicting genre by averaging all node representations. With this, our objective function becomes:
\begin{equation}
    \begin{array}{l}
    L = \min(L_{r}+\lambda L_{C}), \text{where} \\
    L_{r} = \sum_{v}^{\hat{V}}MSE(o_{v}, y_{v})\\   
    L_{c} = -\sum_{g}^{G} L(\sigma(O_g),Y_g)
    \end{array}
\end{equation}
where $\lambda$ is a factor for balancing loss, $\hat{V}$ is the set of musical patterns in which pianoroll latent is masked, $o_{v}$ is the representation output of a model, $y_{v}$ is target pianoroll latent, $G$ is the set of the song structure graph, and $O_g$ is the classification output of the model, and $Y_g$ is the genre label of the graph.

This allows node representation to be obtained not only to predict musical pattern's pianoroll, but also to predict genre features given in music. These node representations are used for another model, by conditioning the generation process. We constructed an Unet\cite{ronneberger2015u} model with node representation as input and piano roll's latent as a target so that the pianoroll of each musical pattern's pianoroll latent can be generated by node representation condition. The whole model architecture is described in Figure \ref{fig:gnn}.

\begin{table}[t]
\centering
        \begin{tabular}{|c|c|c|c|c|c|}
    \hline
    \diagbox[width=10em]{Task}{Metric} & ND & UP & KS & VA & DA \\
    \hline
    ground-truth&20.85&6.23&0.74&94.81&0.81\\
    \hline
    Generation&23.04&6.81&0.70&44.66&0.29\\
    Inpainting&17.60&6.26&0.75&79.10&0.54\\
    Cond. Generation&22.14&6.46&0.72&41.55&0.17\\
    \hline

        \end{tabular}
        \caption{This is the result of the experiment.}
\end{table}
\section{Experiment}\label{sec:typeset_text}

\subsection{Implementation Details}
\subsubsection{Dataset}
We conducted an experiment using cleansed LPD (Lakh Pianoroll Dataset)-17 which is a subset of the Lakh MIDI dataset\cite{dong2018musegan}\cite{raffel2016learning}. It consists multitrack of 17 instruments each, and all track has 4/4 time signature. In addition, we collected 4240 pieces of music, which key information is given in MIDI's internal metadata, and the genre is one of 'country', 'piano', 'rock', 'pop', 'folk', 'electronic', 'rap', 'chill', 'dance', 'jazz', 'rnb', 'reggae', 'house', 'techno', 'trance', 'metal', 'pop\_rock', 'latin', and 'catchy'. We quantized pianoroll to handle minimum length to 48th note. 

\subsubsection{Experiment Details}

During training, we set a loss balancing factor $\lambda$ as 1, and a masking ratio of 30\% for each node in the pianoroll. We used a novel format of pianoroll, CONLON. The CONLON pianoroll incorporates the Velocity channel and Duration channel, so it can contain velocity information that cannot be contained by a classic pianoroll. Also, it contains note data in only onset, so focusing on more crucial features compared to classic pianorolls.

\subsection{Evaluation}

We conducted three experiments to assess the efficacy of our proposed framework. Firstly, we explored inpainting by masking 30\% of a musical pattern to test the model's ability to accurately reconstruct the masked section. Secondly, we embarked on music generation by preserving only musical patterns with a playing time of up to 8 bars, to see how the model generates corresponding patterns. Lastly, in the melody-conditioned music generation task, we hypothesized that the instrument with the highest number of connections via the "Same Homogeneity Edge" serves as the primary instrument. Using a pre-composed melody, we analyzed how the model composes accompaniments for other instruments. To evaluate the performance across these tasks, we employed several metrics that capture musical characteristics. It is crucial to note that these metrics were computed across all masked musical pattern nodes, excluding the drum instrument. Detailed methodologies are further explained in the below.

\begin{description}
    \item[Key Score (KS)] According to 24 basic key scales(C maj, C\#maj, ..., B min), check if notes' pitches are on the scale in the musical pattern.
    $KS = \frac{\text{\# of notes in key}}{\text{\# of notes}}$
    \item[Unique Pitch (UP)] Unique pitches that occur at least once in a musical pattern.
    \item[Note Density (ND)] Count the number of notes in musical pattern.
    \item[Velocity Average (VA)] Average value of velocities.
    \item[Duration Average (DA)] Average value of durations.
\end{description}

\section{Result}

The results of the experiment above are shown in Table 1. For the ground-truth, We averaged metric for randomly selected 1000 pieces of music in dataset. to evaluate the models, we used 43 pieces of music which were not used for training.

In the case of Inpainting, it was confirmed that song metrics did not differ significantly from the ground-truth, but in some instances, the metrics were slightly off. For the generation or melody-conditional generation tasks, the Velocity Average or Duration Average indicators were quite far from ground-truth. As the proportion of provided musical pattern information increased, the metrics approached the ground truth more closely, leading to better performance in the inpainting task. Still, these metrics show that our model can construct a general form of music.

 While our metrics provide objective measurements, our subjective assessment of the generated music sequences is less favorable. While each instrument's individual quality may be satisfactory, the combined multi-track output often displays inconsistent, which undermines the overall performance. However, our results confirm that each musical pattern is generated with consideration of key and instrument information, and graph structure. Specifically, in the case of drum patterns, it is inferred that direct information input could yield high performance, especially when considering the proximity of the generated output to actual usability. Generally, for drum patterns represented in pianoroll, we think it is a phenomenon that occurs because it has distinct characteristics compared to other instruments.
 
Even if our model did not succeed in generating music with complete quality, it was shown that notes could be configured to have a musical distribution similar to the original data using the information provided in generating musical patterns, to construct the entire song structure graph.

\section{Conclusion}

We have proposed a framework for performing a combinatorial music generation task, which involves expressing multitrack symbolic music as a single song structure graph. The framework has shown its ability to generate new music that is similar in form to existing music, as well as perform tasks such as generation and inpainting tasks.

However, the research has several limitations at this point. It is hard to ensure stable generation during the process of connecting various models, like GNN, Autoencoder, and Unet. Furthermore, we believe our proposed framework can be enhanced by utilizing various models such as transformer and diffusion models. Currently, there isn't a process for generating a song structure graph from scratch, so it is difficult to see it as a model that can completely produce entire music. 

Although our song structure analysis considered the novelty, reposition, and homogeneity of each pattern, the study did not include any track-role (e.g., verse or chorus) or explicit expression (e.g., A-B-A'-B). It is hoped that future research will address this limitation based on more detailed analysis and datasets. 

 Our framework well embodies the actual producer's composition method, and has demonstrated certain quality in composition. Beyond this, our framework is expected to have wide applicability not only for music generation but also for other tasks that account for the overall song structure in symbolic music, such as music modeling and harmonization. 

% References should be produced using the bibtex program from suitable
% BiBTeX files (here: strings, refs, manuals). The IEEEbib.bst bibliography
% style file from IEEE produces unsorted bibliography list.
% -------------------------------------------------------------------------
\bibliographystyle{IEEEbib}
\bibliography{strings,refs}

\end{document}